\pdfoutput=1
\documentclass[11pt]{article}
\usepackage[utf8]{inputenc}
\usepackage[T1]{fontenc}
\usepackage[margin=2.5cm]{geometry}
\usepackage{amsmath,amssymb,amsthm}
\usepackage{booktabs}
\usepackage{hyperref}
\usepackage{xurl}
\usepackage{graphicx}
\usepackage{array}

\newtheorem{theorem}{Theorem}
\newtheorem{lemma}{Lemma}

\theoremstyle{definition}

\newcommand{\Tr}{\operatorname{Tr}}

\title{Hidden selection rules and amplitude ceilings in quantum Fourier models}
\author{Pavel Sulimov$^{1,*}$ and Claude Lehmann$^{1}$\\[6pt]
\normalsize $^1$Zurich University of Applied Sciences (ZHAW), Zurich, Switzerland\\
\normalsize $^*$Correspondence: \texttt{suli@zhaw.ch}}
\date{}

\begin{document}
\maketitle

\begin{abstract}
Quantum models with Pauli data encodings compute trigonometric polynomials whatever channels, ancillas and outcome-averaged measurements follow,
so their expressivity depends on the Fourier coefficients a circuit can reach. We identify two kinds of structure that settle this before training.
An antiunitary time-reversal rule and a unitary reflection rule remove frequencies. For gate-level circuits, deciding whether such a rule exists reduces
to a linear system over GF(2), solvable in polynomial time, and the time-reversal rule holds in 17.6\% of 6{,}120 configurations of standard templates.
Single-upload encodings cap the top coefficient at $2^{-m}$ for every channel, and we compute the tight ceiling for every frequency and a depth-independent
re-uploading bound. Frozen before training, these bounds were never violated on a non-linearity benchmark, and they predicted trained
performance on real data. On IBM processors the rule held on a 65-qubit chain, and we switched rule-breaking relaxation noise on and off by design.
\end{abstract}

\section{Introduction}
A parameterised quantum circuit that encodes a scalar $x$ through gates $e^{-ixG/2}$ computes a trigonometric polynomial
$f(x)=\sum_\omega c_\omega e^{i\omega x}$ whose frequencies are set by the eigenvalue differences of the generators~\cite{schuld2021effect}. Composing the
encodings with arbitrary linear maps keeps the output a trigonometric polynomial with the same frequency bound (Lemma~\ref{lem:closure}), and the main
paper~\cite{mainpaper} shows that every finite-shot procedure, with any classical head, is bounded by the spectral range of its encodings. This leaves
open which coefficients a given circuit can actually reach. We show that two kinds of structure decide this, and that both can be computed before
training.

The first kind produces exact zeros. Hidden symmetries of a circuit remove frequencies from its output: we identify an antiunitary (time-reversal)
selection rule and a unitary reflection rule, give a polynomial-time procedure that decides whether a circuit has one, and show that a single gate can
switch the rule off. The second kind produces exact ceilings, which can make even an allowed frequency useless. For single-upload encodings the top
coefficient is at most $2^{-m}$ whatever channel follows, and we compute the tight ceiling for every frequency. Together the two turn a benchmark into a
test of predictions. Every failure of the six ans\"atze of the QNLEP benchmark~\cite{qnlep2026} is either structurally impossible or capped by the
ceiling, and we froze these predictions before the training runs that test them.

Earlier work computes exact spectra and coefficients for Clifford-plus-rotation circuits~\cite{nemkov2023fourier,wiedmann2025fourier}, bounds the
largest frequency from the encoding~\cite{schuld2021effect,holzer2026spectral}, and studies coefficient magnitudes for typical (Haar or 2-design)
parameters~\cite{mhiri2025constrained,barthe2024gradients}. Wiedmann et al.\ report frequencies that vanish for reasons that ``cannot be explained by the
heuristics''~\cite{wiedmann2025fourier}, and symmetries of the kind we describe are one such reason. Group-invariant and equivariant QML imposes
symmetries of the data by design~\cite{larocca2022group,meyer2023exploiting}, whereas the symmetries studied here are unintended properties of a circuit
that remove expressivity. Noise is usually described as contracting Fourier coefficients~\cite{fontana2022spectral,franz2025outoftune}; we show that
non-unital noise can also \emph{restore} frequencies that the noiseless circuit provably cannot express. Re-uploading models are known to suppress high
frequencies on average over Haar-random layers~\cite{barthe2024gradients}; our ceilings hold in the worst case, over all parameters and all channels.

\section{Results}
\subsection{Channels do not change the function class}
\begin{lemma}[Channel closure~\cite{mainpaper}]\label{lem:closure}
Let $x$ enter only through encoding gates $U_j(x)=e^{-ixG_j/2}$ whose generators have integer eigenvalue half-differences $(\lambda-\lambda')/2$ and spectral range
$\Delta_j=\lambda_{\max}(G_j)-\lambda_{\min}(G_j)$. Interleave them with arbitrary $x$-independent unitaries, CPTP maps and instruments, where measurement outcomes may control
later operations, including whether later encodings are applied. Then the outcome-averaged model $f(x)=\Tr[O\rho_{\mathrm{out}}(x)]$ is a trigonometric polynomial with
$|\omega|\le\max_{\mathrm{branches}}\sum_{j}\Delta_j/2$. For Pauli generators ($\Delta_j=2$) this is $N_{\max}$, the largest number of encodings on any outcome branch.
\end{lemma}
The proof and the numerical checks are in~\cite{mainpaper}. Within this class, the remaining question is which coefficients a given circuit can reach.

\subsection{A time-reversal selection rule}
\begin{theorem}[Time-reversal selection rule]\label{thm:tr}
Let the encoded qubits $\mathcal E$ be prepared once as $R_Y(x)|0\rangle$ and the rest of the register in $\rho_R$. Let $\Phi$ be any CPTP map and $f(x)=\Tr[O\,\Phi(\rho_x)]$. If the
antiunitary $A=(\bigotimes_{j\in\mathcal E}Y_j\otimes V_R)K$ satisfies $A\rho_RA^{-1}=\rho_R$, $\Phi(A\sigma A^{-1})=A\Phi(\sigma)A^{-1}$ for all $\sigma$, and $A^{-1}OA=sO$ with
$s=\pm1$, then $f(x+\pi)=s f(x)$. Hence $c_\omega=0$ unless $(-1)^\omega=s$.
\end{theorem}
Spin time reversal $YK$ commutes with every single-qubit rotation, so trainable single-qubit layers never break the rule; the entangling structure and
the noise can. A CNOT from an encoded qubit into an ancilla on which $A$ acts as $K$ maps $Y_c$ to $Y_cX_t$, so a star of $n$ such CNOTs is covariant if
and only if $n$ is even. This is why the ancilla-tracing ansatz of the QNLEP benchmark expresses only odd frequencies for every even $n$ and every depth,
and why its $k=2$ rung collapsed at $(n,L)=(4,4)$~\cite{qnlep2026}. Depolarizing and dephasing channels are covariant and preserve the rule. Amplitude
damping is not covariant, because $YK$ maps relaxation toward $|0\rangle$ onto relaxation toward $|1\rangle$. A non-unital channel can therefore add
frequencies that the noiseless model cannot express, while staying inside the encoding spectrum.

\paragraph{Reflection rule.} A unitary symmetry acts differently. Since $X\,R_Y(x)|0\rangle=\pm R_Y(\pi-x)|0\rangle$, an $X$-type frame on the encoded
qubits that commutes with the rest of the circuit and anticommutes with $O$ gives $f(\pi-x)=-f(x)$. This forces $c_0=0$ and fixes the phases
$c_{-k}=-(-1)^kc_k$, but removes no other frequency. It explains the only circuit in our 144-circuit design census on which the Pauli-path bound
(Methods) was not exact, a trainable-$R_X$ system with an ancilla $R_Z$ whose support is $\{1,2,4\}$ with $c_0\equiv0$ (verified to $2\times10^{-16}$).

\paragraph{Deciding the rule.} For gate-level circuits the search for a frame reduces to a linear system over $\mathrm{GF}(2)$, so whether a rule exists
can be decided in polynomial time; a circuit with 1{,}025 qubits and 31{,}744 gates takes 51\,ms (validation in Methods, ``The GF(2) frame checker and
the census''). On a census of 6{,}120 configurations of standard templates and libraries the time-reversal rule holds in 17.6\% of cases (44\% of ancilla
designs and 30\% of template configurations), and a Pauli-path support bound (Methods) is exact on 203 of 204 benchmark circuits, the exception being the
reflection rule. The rule is therefore not a curiosity of one ansatz.

\subsection{An exact amplitude ceiling}
\begin{theorem}[Amplitude ceiling]\label{thm:ceiling}
Let $m$ qubits each carry one encoding $R_Y(x)$ on $|0\rangle$, followed by any CPTP map (with ancillas, noise and outcome-averaged measurements) and a readout with $\|O\|\le1$.
Write $|\psi(x)\rangle\langle\psi(x)|=\sum_kM_ke^{ikx}$. Then
\[
\max|c_k| = c^*(m,k) := \max_{\varphi}\big\|\tfrac12(e^{i\varphi}M_k+e^{-i\varphi}M_k^\dagger)\big\|_1,
\]
and the maximum is attained. In particular $c^*(m,m)=2^{-m}$.
\end{theorem}
Because $|\psi(x)\rangle$ is a product of identical qubit states, it lies in the symmetric subspace, and computing $c^*$ is an $(m{+}1)$-dimensional
eigenvalue problem (Table~\ref{tab:ceiling}). Fitting $\sin(kx)$ with an amplitude of at most $a=2c^*$ gives $R^2\le1-(1-a)^2$. With eight single-upload
qubits the $k=8$ component exists but caps the fit at $R^2\le0.016$, while $k=4$ caps it at $0.87$. Single-upload models thus carry high frequencies only
with exponentially small weight, for every channel. The ceiling is the Fourier form of the separable limit: a single upload acts on a product state, the
master bound of~\cite{mainpaper} bounds every coefficient by $e^{-k^2/2m}$, and an entangled input reaches $\tfrac12$ at $k=m$. It is a deterministic,
worst-case counterpart of the average-case results for Haar-random layers~\cite{barthe2024gradients,mhiri2025constrained}. The ceiling also explains
$R^2$ plateaus that appear in training regardless of budget: $0.750$, $0.437$ and $0.234$ are the ceilings for $k=m=2$, 3 and 4, and a tenfold training
budget does not move them.

\begin{table}[h]\centering
\caption{Exact amplitude ceiling $c^*(m,k)$ for an $m$-qubit single upload. The diagonal is $2^{-m}$.}\label{tab:ceiling}
\begin{tabular}{lccccccc}\toprule
$m\backslash k$ & 1 & 2 & 3 & 4 & 5 & 6 & 8\\\midrule
2 & 0.500 & 0.250 &&&&&\\
3 & 0.573 & 0.433 & 0.125 &&&&\\
4 & 0.576 & 0.467 & 0.250 & 0.0625 &&&\\
6 & 0.599 & 0.506 & 0.395 & 0.215 & 0.077 & 0.0156 &\\
8 & 0.610 & 0.540 & 0.446 & 0.321 & 0.175 & 0.073 & 0.0039\\
12 & 0.619 & 0.570 & 0.500 & 0.416 & 0.310 & 0.200 & 0.052\\\bottomrule
\end{tabular}
\end{table}

\begin{figure}[t]\centering
\includegraphics[width=\textwidth]{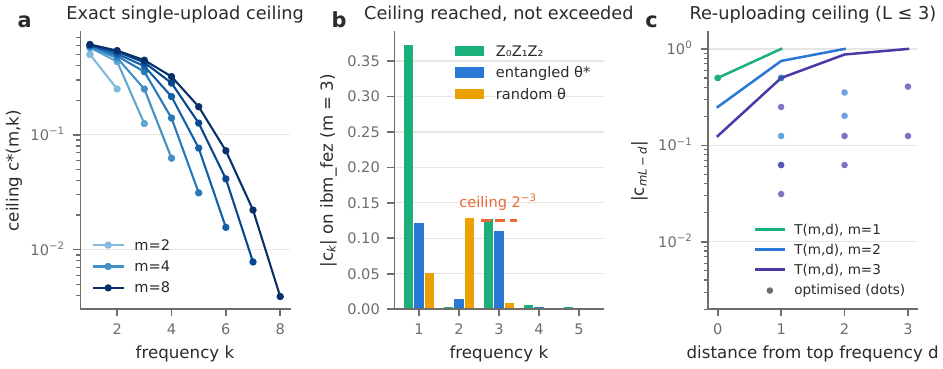}
\caption{\textbf{Amplitude ceilings.} \textbf{a}, exact single-upload ceiling $c^*(m,k)$; the top coefficient is $2^{-m}$. \textbf{b}, $|c_k|$ measured on ibm\_fez for three qubits: the product readout $Z_0Z_1Z_2$ ($\cos^3x$), an entangled circuit whose parameters maximise $|c_3|$ of $\langle Z_0\rangle$ (88\% of $2^{-3}$), and a random circuit; no arm exceeds the ceiling. \textbf{c}, re-uploading ceiling $T(m,d)$ (lines) and the largest coefficients found by optimisation for $m\le3$, $L\le3$ (dots).}\label{fig:ceiling}
\end{figure}

\paragraph{Re-uploading.} For $m$ qubits re-uploaded $L$ times, with any channels in between,
\[
|c_{mL-d}|\le T(m,d)=\Pr\big[\mathrm{Bin}(m,\tfrac12)\ge m-d\big]=2^{-m}\sum_{i\le d}\binom mi ,
\]
independent of $L$ (Methods). At $d=0$ this is $2^{-m}$, so the frequencies at the top of a re-uploading spectrum stay exponentially small in the
width of the first upload layer, however deep the model. The bound is tight at $d=0$ and loose below, and numerical maximisation for $m\le3$, $L\le3$
found no violation (Fig.~\ref{fig:ceiling}c).

\subsection{Structure predicts the benchmark}
\begin{figure}[t]\centering
\includegraphics[width=\textwidth]{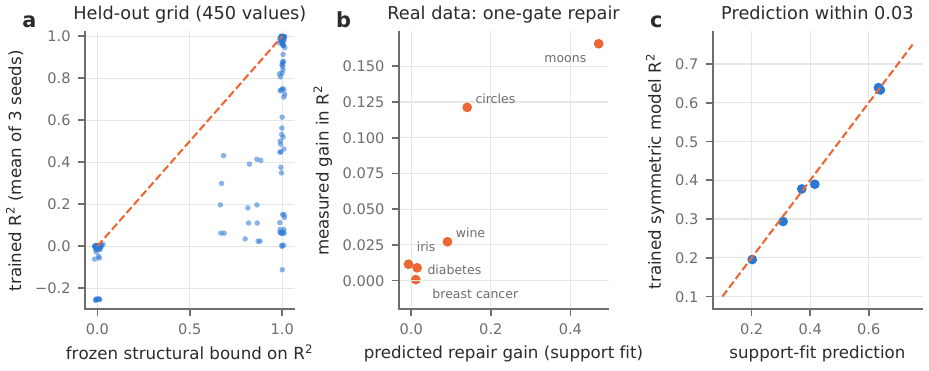}
\caption{\textbf{Structure predicts trained performance.} \textbf{a}, held-out grid: trained $R^2$ of every rung against the structural bound frozen before training (450 values, no violation). \textbf{b}, six real datasets: the gain from the one-gate repair against the gain predicted by a classical fit on the Fourier supports (Spearman 0.83). \textbf{c}, the same classical fit predicts the trained symmetric model within 0.03.}\label{fig:prediction}
\end{figure}

Support and ceiling together bound every rung of the QNLEP non-linear-capacity benchmark~\cite{qnlep2026} before training. We froze the bounds before
training on 72 cells and on a held-out grid of five operating points with up to 8 qubits. No forbidden rung was fitted (0 of 965 and 0 of 183), and no
trained value exceeded its bound (0 of 450). The bounds rank methods only partly (Kendall $\tau=0.44$ on the held-out grid, below the pre-registered
0.6). Structure decides what \emph{can} be fitted and how well \emph{at best}, and training decides the rest. Adding one $R_Z$ on the ancilla per layer
breaks the rule without new parameters and triples the benchmark score at $(4,4)$. On real two-feature data, a classical fit on each model's Fourier
support, frozen before training, predicted the trained model's test $R^2$ within 0.026 on six datasets. In a fresh replication on 20 tasks it did so
within 0.15 in all 60 cells; the predicted sign of the repair's effect held in 19 of 19 cases, and no effect appeared in 20 of 20 cases where none was
predicted (Fig.~\ref{fig:prediction}; details in Methods).

\subsection{Hardware}
\begin{figure}[t]\centering
\includegraphics[width=\textwidth]{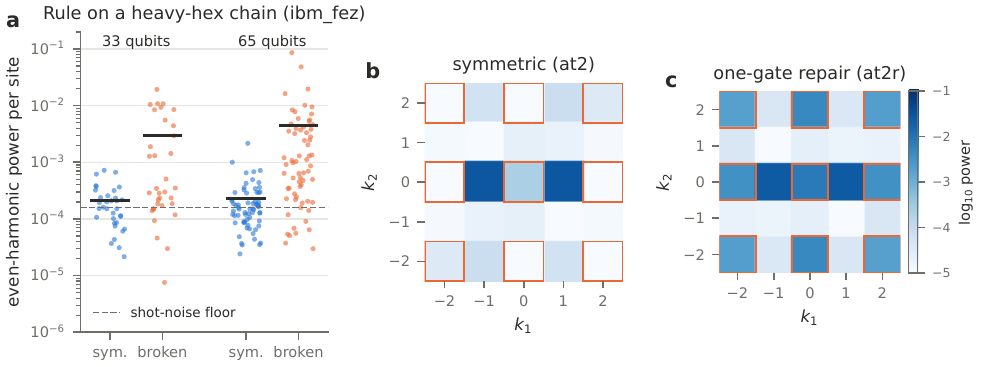}
\caption{\textbf{The selection rule at scale and in two dimensions.} \textbf{a}, even-harmonic power of every system-qubit readout of a heavy-hex ancilla chain on ibm\_fez at 33 and 65 qubits, for the symmetric circuit and a control with one $R_Z$ per ancilla (bars: means). \textbf{b, c}, 2-D Fourier power of the ancilla-tracing circuit and its one-gate repair on ibm\_fez; outlined cells (even $k_1$, even $k_2$) are forbidden by the rule.}\label{fig:scale}
\end{figure}

\begin{figure}[t]\centering
\includegraphics[width=\textwidth]{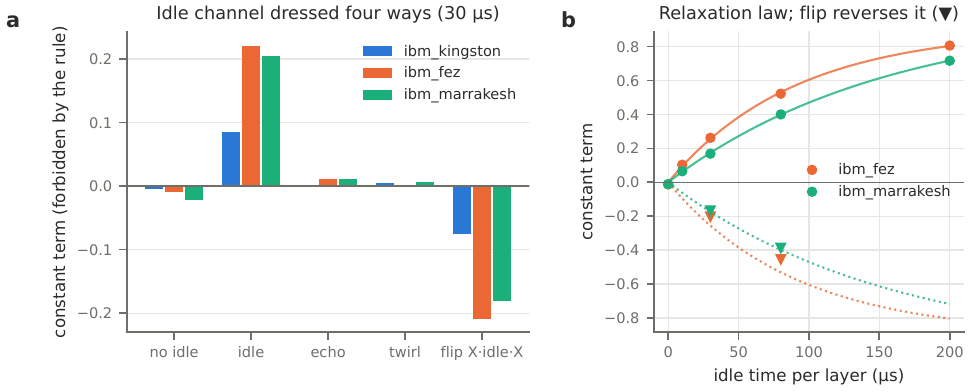}
\caption{\textbf{Rule-breaking noise switched by design.} \textbf{a}, constant term (forbidden by the rule) with 30\,$\mu$s idles left plain, echoed, Pauli-twirled, or run in the flipped frame, on three devices. \textbf{b}, the offset against idle time follows a relaxation law (lines: fits), and the frame flip reverses it at every dose (triangles; dotted: mirrored fits).}\label{fig:noise}
\end{figure}

Every run was pre-registered with thresholds and a same-day noise prediction before submission (Table~\ref{tab:hw}; Methods describes the commit
record), and we report the misses alongside the passes. The selection rule held on hardware wherever its symmetry did. Forbidden harmonics stayed at the
shot-noise floor for a five-qubit ancilla circuit and for a heavy-hex ancilla chain at 33 and 65 qubits, where the even power per site was $2.1\times10^{-4}$ and
$2.3\times10^{-4}$, against 14 and 19 times more for a control (Fig.~\ref{fig:scale}a). They also stayed at the floor for the two-dimensional rule and
its one-gate repair (Fig.~\ref{fig:scale}b,c), and for a hidden rule in a one-layer StronglyEntanglingLayers template, whose forbidden half stayed at 1.2
to 3 times the floor while two-layer controls showed it 97 and 147 times higher. Rule-breaking noise could be switched on and off by design
(Fig.~\ref{fig:noise}). Plain idles broke the rule through relaxation, an echo or a Pauli twirl restored it, running the idle in the bit-flipped frame
reversed the offset, and the offset followed the relaxation law with the calibrated time constant; one device also revealed a coherent $Z$ phase. A
three-qubit circuit optimised for $|c_3|$ reached 88\% of the ceiling $2^{-3}$, and no arm exceeded the ceiling, as the theorem requires for every
channel, noise included (Fig.~\ref{fig:ceiling}b). Methods gives the details, including the misses.

\section{Discussion}
The common picture of ``sources of quantum non-linearity'' mixes up the function class with its coefficients. Ancillas, partial traces, measurements
averaged over outcomes, reservoirs and noise all produce the same class of functions, and differ in which frequencies they leave and how much weight
those frequencies can carry. Both properties follow from the circuit structure and can be computed in polynomial time for gate-level circuits, so a
benchmark can report them before any training, as a ``stage zero''. How much non-linearity a model can buy at all, and at what price in encodings and
shots, is the subject of the main paper~\cite{mainpaper}.

On hardware the selection rule doubles as a noise witness. A non-covariant channel, such as relaxation or a coherent phase on an ancilla, shows up as
forbidden harmonics or as an offset that the noiseless circuit cannot produce, and dressing the idle channel with an echo, a twirl or a frame flip tells
the mechanisms apart. The design rule behind our 65-qubit test, that every ancilla collects an even number of system CNOTs per layer, gives
hardware-native circuits with a certified zero at any size.

These results have limits. The selection rules are proven for single uploads of $R_Y$ or $R_X$ encodings and for the frame class. The checker is sound
and complete for that class, but other symmetries exist; the reflection rule is one we found by examining the single failure of the path bound. The
ceilings are tight over all channels rather than over a given ansatz, so they are upper envelopes, not predictions of trained performance, and on real
data they predict well only when the structure is the binding constraint. Four of the eleven experiments missed at least one pre-registered criterion: a
replication with an ill-conditioned criterion, the 9--17-qubit run because of a relaxation offset, its decoupled repeat at two of three sizes by at most
0.002, and one zero-delay criterion by 0.001 on one of three devices. All runs used IBM Heron processors.

\section{Methods}
\subsection*{Proof of Theorem~\ref{thm:tr}}

$YK\,R_Y(x)|0\rangle=i\,R_Y(x+\pi)|0\rangle$, so $A\rho_xA^{-1}=\rho_{x+\pi}$. Then
$f(x+\pi)=\Tr[O\,A\Phi(\rho_x)A^{-1}]=\overline{\Tr[(A^{-1}OA)\Phi(\rho_x)]}=s f(x)$, using that $f$ is real.

\subsection*{Proof of Theorem~\ref{thm:ceiling}}

$c_k=\Tr[HM_k]$ with $H=\Phi^\dagger(O)$ a Hermitian contraction, because $\Phi^\dagger$ is unital and positive. For every phase, $\mathrm{Re}(e^{i\varphi}c_k)\le\|\mathrm{Herm}(e^{i\varphi}M_k)\|_1$
by H\"older, with equality for $H$ the sign of that matrix. Any Hermitian contraction is $\Phi^\dagger(Z)$ for the two-outcome POVM $\{(I\pm H)/2\}$ on a dilation, so the bound
is attained. For $k=m$, $M_m=\bigotimes_j(Z_j-iX_j)/4$ has trace norm $2^{-m}$ and is already of the required form.

\subsection*{Proof of the re-uploading ceiling}
Track, for each encoding $e^{-ixY/2}=e^{-ix/2}P_++e^{ix/2}P_-$, which branch a path takes. Dilate the intermediate channels (Stinespring) and add a counter register that each encoding increments on its $P_+$ branch, a controlled
increment that keeps the evolution an isometry; the final state is $\sum_j|\phi_j\rangle|j\rangle$, with orthogonal components $|\phi_j\rangle$ labelled by the number $j$ of $+$ branches, and
$c_k=\sum_j\langle\phi_{j-k}|O|\phi_j\rangle$. Only the first upload acts on $|0\rangle$, where each branch has probability $\tfrac12$; later uploads act on arbitrary states.
A component whose path takes at least $mL-d$ plus-branches must have at least $m-d$ of them in the first layer, so the components with $j\ge mL-d$ have total weight at most $T(m,d)=\Pr[\mathrm{Bin}(m,\tfrac12)\ge m-d]$. Their partners in $c_{mL-d}$ have $j-(mL-d)\le d$ plus-branches, hence at most $d$ in the first layer, and by symmetry of the first-layer branches their total weight obeys the same bound.
Cauchy--Schwarz over the two sets of components, with $\|O\|\le1$, then gives $\sqrt{T(m,d)}\sqrt{T(m,d)}$:
\[
|c_{mL-d}|\le T(m,d)=2^{-m}\sum_{i\le d}\binom mi
\]
for $m$ qubits re-uploaded $L$ times with any channels in between. For $d=0$ this is $2^{-m}$, the single-upload top ceiling, \emph{independent of $L$}:
the frequencies near the top of a re-uploading spectrum stay exponentially small in the width of the first upload layer, however deep the model.
The bound is tight at $d=0$ for every $L$ (the product readout $Z^{\otimes m}$ without entanglers gives $\cos^m(Lx)$, whose top coefficient is $2^{-m}$) and loose below the top
(for $L=1$ it exceeds the exact $c^*(m,k)$). Maximising $|c_k|$ over the parameters of data re-uploading circuits with $m\le3$, $L\le3$ gave no violation in 24 cases.

\subsection*{Pauli-path support bound}
Propagating the readout backwards and counting, for each Pauli string, the number $b$ of encodings it branched on gives a sound bound:
$\mathrm{supp}(f)\subseteq\bigcup\{\omega:|\omega|\le b,\ \omega\equiv b\ (\mathrm{mod}\ 2)\}$ over strings in $\{I,Z\}^N$ that survive to $|0\cdots0\rangle$. It ignores
cancellations, needs no coefficients and runs as a dynamic program over distinct Pauli strings, unlike the path-enumeration trees used for exact
coefficients~\cite{nemkov2023fourier,wiedmann2025fourier}. On the 144-circuit design census and the QNLEP registry on a $4\times3$ grid of $(n,L)$ the bound is sound in 204 of 204
circuits and exact in 203. The remaining case is the reflection rule above, so the two together explain every support we measured.
With one branch counter per input variable the bound extends to several inputs; it reproduces the exact 2-D supports of all three models of the breadth test
(33, 6 and 25 frequency pairs, including the per-variable parity rule). Its cost is set by the readout's light cone, not by the circuit size. For data re-uploading with a
CNOT chain it stays at 239 Pauli strings for $L=4$ from 4 to 32 qubits, but for scrambling layouts it grows as $\sim4^n$ (about $10^6$ strings at $n=10$). The path bound is
therefore a small-to-medium-$n$ tool; the frame checker is the one that scales.

\subsection*{The GF(2) frame checker and the census}
For gate-level circuits (encodings $R_P(x)$, trainable rotations $R_P(\theta)$, Clifford gates, Pauli-string readout) we search for a frame $F=QK^k$, $Q$ a Pauli string,
with $F|0\cdots0\rangle\propto|0\cdots0\rangle$, that is carried through the circuit so that $U(x+\pi)|0\rangle\propto F_{\mathrm{end}}U(x)|0\rangle$. The initial frame is
$Z^{z_0}$ for an unknown $z_0\in\mathrm{GF}(2)^N$. Every later frame bit is an affine function of $z_0$: Clifford conjugation is linear on the symplectic vector and an encoding
multiplies the frame by a fixed Pauli. Every gate condition (the frame must commute, or for real rotations under $k=1$ anticommute, with the rotation generator) is one
affine equation. A frame exists if and only if the linear system is consistent, which Gaussian elimination decides in $O(GN^2/w)$ word operations for $G$ gates and $N$ qubits.
The observable sign is $s=(-1)^{a+k\,n_Y}$, where $a$ counts anticommuting factors of $O$ with the final frame and $n_Y$ counts $Y$ factors of $O$ ($KYK=-Y$).
If the solution space admits both signs, $f\equiv0$.

The procedure is complete for the frame class and sound. It agrees with exhaustive frame enumeration on 20{,}000 of 20{,}000 random circuits. Against exact simulation of
3{,}000 random circuits with $R_X/R_Y/R_Z$ encodings and random Pauli readouts, all 1{,}208 claimed rules hold with error at most $3\times10^{-15}$, and all 540 circuits
with both signs admissible are identically zero. The reflection rule is the same linear system with every encoding condition flipped (the frame must map
$R_P(x)$ to $R_P(-x)$); on the same 3{,}000 circuits its 896 claims hold with error at most $9\times10^{-16}$. The checker decides the ancilla-tracing family at 1{,}025 qubits and 31{,}744 gates in 51\,ms, where enumeration of $2^{N+1}$ frames
is impossible; every claim up to 17 qubits matches exact simulation to $7\times10^{-16}$.

\paragraph{Census.} We ran the checker on 6{,}120 configurations: the 19 circuits of Sim et al.~\cite{sim2019expressibility}, six standard templates, a
QCNN and three ancilla designs, with $n=4$--6, $L=1$--3, $R_X$, $R_Y$ or $H{+}R_Z$ encodings, single or repeated upload, and four readouts, each compared with its exact
spectrum. The time-reversal rule holds in 1{,}080 configurations (17.6\%), all confirmed and none violated; in 942 of them $f\not\equiv0$. It is common in ancilla designs (44\%)
and in the BasicEntangler, StronglyEntangling, EfficientSU2 and RealAmplitudes templates (30\% of template configurations), rare in the Sim et al.\ library (9\%, non-trivially
only in circuits 1, 2 and 10), and more frequent for single uploads (22\%) than for repeated ones (14\%). Every parity-restricted spectrum in the census has an identified cause:
the time-reversal rule (1{,}080), the path parity of the Pauli-path bound (480), a path-bound zero (178) or the $\pi-x$ reflection (48). The $x\mapsto-x$ reflection, the same linear
system without the frame update at encodings, explains 126 of the 176 remaining cases in which only $c_0$ vanishes; 50 cases in two CZ/Hadamard-based families stay unexplained
and mark the limit of the frame class. All three rule types pass the random-circuit validation (1{,}208, 896 and 906 claims, no violation).

\subsection*{Statistics}
Numbers after $\pm$ are standard errors (shot noise for hardware values) unless stated otherwise. Zero counts are reported with the exact one-sided 95\% upper bound on the underlying rate, $1-0.05^{1/n}$: no forbidden rung fitted in 965 trials ($<0.31\%$), no forbidden 2-D
target in 141 ($<2.1\%$), no violation of the frozen bound in 450 ($<0.66\%$) and no impossible real-data fit in 450 ($<0.66\%$).
Rank correlations carry bootstrap 95\% intervals over cells or datasets: held-out Kendall $\tau=0.44$ $[0.15,0.68]$ (30 cells); first-feature real-data $\tau=0.08$
$[-0.06,0.24]$; repair-gain Spearman $0.83$ $[0.03,1.0]$, wide because it rests on six datasets. Per-dataset repair gains and surrogate deviations are bootstrapped over seeds.
Hardware errors are shot-noise standard errors from the estimator or bootstraps over sites (65-qubit chain: contrast 14 $[7,23]$ at 33 qubits,
19 $[9,37]$ at 65 qubits). The relaxation-law time constants are $\tau=90$ $[87,93]\,\mu$s (fez) and $156$ $[147,167]\,\mu$s (marrakesh).

\subsection*{Structure predicts the benchmark (details)}
The QNLEP non-linear capacity (NLC) fits $\sin(kx)$ for $k\in\{1,2,4,8,16\}$ and integrates the fit quality over $\log k$~\cite{qnlep2026}. We pre-registered
the predictions below before each training run, using the benchmark's own training recipe.

\paragraph{Forbidden rungs.} On a grid of $n\in\{2,\dots,5\}$, $L\in\{2,3,4\}$ and all six ans\"atze (72 cells, 5 seeds), no rung outside the support was ever fitted:
0 of 965 forbidden cases reached $R^2>0.05$, and 88\% of allowed rungs did. In two dimensions (three models, 24 targets $\sin(k_1x_1+k_2x_2)$, three seeds) all
141 forbidden targets failed and all 75 allowed targets were fitted. The ancilla-tracing model obeys a per-variable parity rule that we derived before computing:
odd $k_1$, even $k_2$.

\paragraph{Ceiling-based bounds.} Combining support and ceiling gives an upper bound on every rung's $R^2$. On the 72-cell data it was never exceeded:
0 violations in 1{,}800 trained values (a post-hoc test). We then froze the bound on a held-out grid of five operating points with up to 8 qubits and
6 layers, $(n,L)\in\{(6,2),(6,4),(7,3),(8,3),(8,6)\}$, before training it with three seeds. The bound held in all 450 trained values (Fig.~\ref{fig:prediction}a), no forbidden rung was fitted (0 of 183),
and the method with the largest bound was the measured leader in 4 of 5 operating points. The sharpest prediction concerned allowed but capped rungs: at $n=8$ the $k=8$ rung lies
inside the support of the strongly entangling, QCNN and reservoir models, yet the ceiling allows $R^2\le0.016$; training reached $0.01$.

\paragraph{Ranking.} As a ranking predictor of NLC, the structural bound reaches Kendall $\tau=0.47$ on the original grid and
$\tau=0.44$ (95\% CI 0.15--0.68) on the held-out grid, below the pre-registered threshold of 0.6. The interval excludes zero, but the bound does not rank methods: the quantum reservoir, which trains only its readout, sits far below its structural bound. Structure decides whether a frequency \emph{can} be fitted and how well \emph{at best}; how close training gets is a separate question.

\paragraph{One-gate repair.} Adding one $R_Z$ on the ancilla per layer breaks the antiunitary covariance without adding parameters. At $(4,4)$ the support grows
from $\{1,3\}$ to $\{1,2,3,4\}$, the $\sin 2x$ fit goes from $R^2=0.00$ to $0.80$, and NLC triples from $0.125$ to $0.382$. The price is trainability: the plateau
indicator BPPI rises from $0.094$ to $0.218$.

\paragraph{Real data.} On six real two-feature datasets (circles, moons, iris, breast cancer, wine, diabetes), with $x_1$ on two qubits and $x_2$ on two others, the
ancilla-tracing model has only three Fourier components (odd $k_1$, even $k_2$), and the one-gate repair adds the even-$k_1$ ones. We fitted each model's Fourier
support classically (ridge regression on the training split) and froze the resulting predictions before training any quantum model. The classical fit predicted the
trained symmetric model's test $R^2$ within 0.026 on all six datasets. The repair helped where the data need the forbidden components (circles $+0.12$, moons $+0.17$) and did
nothing where they do not (iris, breast cancer, diabetes: $\le0.012$), with Spearman $0.83$ (95\% CI $0.03$--$1.0$, six datasets) between predicted and measured gains. A second test on 20 tasks gave Spearman $0.72$ (95\% CI $0.38$--$0.92$) but failed one hypothesis, because its frozen rule put two tasks where the repair was predicted to hurt, and did, into a no-effect bin. A third, on 20 further tasks with three data splits each and a sign-aware rule fixed in advance, passed every hypothesis: the predicted sign held in 19 of 19 cells where an effect was predicted, no effect appeared in 20 of 20 where none was, Spearman $0.75$ (95\% CI $0.55$--$0.89$), and the support fit predicted the trained model within $0.15$ in all 60 cells; all four pre-registered hypotheses passed (Fig.~\ref{fig:prediction}b,c).
An earlier pre-registered test that fed only the first feature to every model, with the benchmark's short training budget, found the structural bounds sound (0 impossible fits
in 450 runs) but not predictive of the ranking (Kendall $\tau=0.08$): there, training rather than structure was the binding constraint. Structure predicts real-data
performance when the structure binds.

\subsection*{Selection-rule hardware (details)}
Every run was pre-registered with thresholds and a same-day noise prediction before submission (IBM Heron: ibm\_marrakesh, ibm\_kingston, ibm\_fez;
eleven experiments, 1{,}535 QPU seconds in total). We report the failures alongside the passes.

\paragraph{Five-qubit ancilla circuit (runs 1, 2a and Q1).} For ancilla tracing with $n=4$ the constant term was $-0.002\pm0.004$ and the even
harmonic power sat at the shot-noise floor, while the $n=3$ control without symmetry showed ten times more. Idle time produced a positive offset that grew with the delay
($+0.062$ at $10\,\mu$s, $+0.126$ at $30\,\mu$s, within $1.7\times$ and $1.14\times$ of the calibrated prediction), and a two-pulse echo removed it ($+0.004$). A single $50\,\mu$s
delay re-created forbidden $k=2,4$ harmonics 31 times their zero-delay level, which the echo suppressed twentyfold: noise can add frequencies the noiseless model provably lacks.
A replication of the first run on two more devices failed its pre-registered zero-delay criterion. The criterion divided floor-level even power by the odd power of a parameter
draw that had almost no signal. The absolute even power was at the floor on all three devices, and the experiments below use absolute criteria.

\paragraph{Idle-noise contrast on three devices (Q4 and Q5).} With the same $30\,\mu$s idles we dressed the idle channel four ways
(Fig.~\ref{fig:noise}a). Plain idles break the rule (offset $+0.086$, $+0.221$, $+0.204$ on kingston, fez, marrakesh). An echo or a balanced Pauli twirl, which make the idle channel
unital and covariant, restore it ($|\mathrm{dc}|\le0.012$). Running the idle in the bit-flipped frame, $X$--delay--$X$, reverses the offset at the same size
($-0.075$, $-0.208$, $-0.180$; ratios 0.88, 0.94, 0.88), which identifies the non-unital part as relaxation. On fez all four arms match the calibrated amplitude-damping
model within 4\%. Kingston revealed a second, coherent breaking channel: even harmonics 15 times the floor that survive the frame flip but not the echo or the twirl, as the
theory predicts for a coherent $Z$ phase on the ancilla. All criteria passed on kingston and fez; on marrakesh one zero-delay criterion missed by 0.001. The offset follows the
relaxation law $A(1-e^{-t/\tau})$ from 0 to $200\,\mu$s (Fig.~\ref{fig:noise}b) with $\tau=90$ and $156\,\mu$s on fez and marrakesh (within 0.65--0.80 of the calibrated value), and the frame flip
reverses it at every dose.

\paragraph{Scale from 9 to 65 qubits (runs 3, 3b and Q3).} The star-shaped ancilla circuit kept the forbidden oscillating harmonics at the noise level up to 17 qubits and 202 two-qubit gates, with a
constant relaxation offset that dynamical decoupling of the idle windows removed by 61--93\%. The star does not route on heavy-hex at larger sizes (depth 2{,}380 at 65 qubits).
The frame argument, however, needs only that each ancilla collects an even number of system CNOTs per layer. A chain $s_0-a_0-s_1-\cdots$ in which every ancilla receives
one CNOT from each neighbour has the same rule for every length, with nearest-neighbour gates only. On ibm\_fez at 33 and 65 qubits (depth 31 and 36, 96 and 192 CNOTs, all system
readouts measured jointly) the even power per site was $2.1\times10^{-4}$ and $2.3\times10^{-4}$, at the shot-noise floor, while the control with an $R_Z$ on each ancilla
showed 14 and 19 times more (95\% CI 7--23 and 9--37), at 74--77\% of the exact ideal (Fig.~\ref{fig:scale}a). The odd signal kept 95--120\% of its ideal power. The device at this size cannot be simulated, but the rule fixes
the answer, and nothing in the symmetric chain grew from 33 to 65 qubits.

\paragraph{Two-dimensional rule and its repair (Q7 and Q8).} For the 2-D ancilla-tracing circuit of the real-data test, the forbidden even-$k_1$ components measured $3.0\times10^{-4}$
(three times the floor); the one-gate repair raised them to $2.9\times10^{-2}$ (81\% of the prediction), while the even-$k_2$ rule, which the repair does not touch, held in both (Fig.~\ref{fig:scale}b,c).

\paragraph{Hidden rule in a standard template (Q10).} The census predicts that a one-layer StronglyEntanglingLayers model with a single $R_Y$ upload expresses only odd
frequencies at four qubits and only even ones at five, and that a second layer removes the rule. Nothing in the template's definition suggests this. On ibm\_fez the forbidden
half of the spectrum stayed at 1.2--3 times the shot-noise floor for the one-layer circuits, while the two-layer controls showed it 97 and 147 times higher, at 81--83\% of the
exact prediction (all criteria passed). Hidden selection rules are a property of ordinary circuits, and they survive hardware noise.

\paragraph{Ceiling saturation (Q6).} A three-qubit entangled circuit whose parameters maximise $|c_3|$ of a single-qubit readout reached $0.110\pm0.004$ on ibm\_fez, 88\% of the
ceiling $2^{-3}$; the product readout $\cos^3x$ gave $0.127\pm0.004$; no arm exceeded the ceiling beyond noise (Fig.~\ref{fig:ceiling}b).

\subsection*{Pre-registration and all hardware experiments}
Each experiment was frozen in a dated note with thresholds, a pass rule and its interpretation fixed in advance, and committed with the same-day prediction from the
device's calibrated Aer noise model before submission. For runs 1, 2a and 3 (September 2026) the notes and prediction files were written before the jobs (the prediction
files carry creation times minutes before each job) but entered version control only the following morning. From run 3b onward every
pre-registration and prediction was committed before the job, and no criterion was changed after a job was submitted. Of the 11 runs, 7 passed every gating
criterion (Table~\ref{tab:hw}).
\begin{table}[h]\centering\small
\caption{Pre-registered hardware experiments on selection rules and ceilings (IBM Heron, second open-plan key unless noted). QPU seconds as charged. The other sixteen runs of the programme are in~\cite{mainpaper}.}\label{tab:hw}
\begin{tabular}{>{\raggedright\arraybackslash}p{4.6cm}>{\raggedright\arraybackslash}p{2.4cm}rl>{\raggedright\arraybackslash}p{3.4cm}}\toprule
Experiment & device & QPU s & test & outcome\\\midrule
1. Rule, T1 dose, echo (5 qubits) & marrakesh & 95 & P1--P4 & pass\\
2a. Non-unital noise re-creates harmonics & kingston & 94 & 3 criteria & pass\\
3. Rule at 9--17 qubits & kingston & 88 & T1--T4 & fail (relaxation offset)\\
3b. Decoupling vs offset, 9--17 qubits & kingston & 88 & D0--D3 & pass at 17; miss $\le0.002$ at 9, 13\\
Q1. Replication of 1 & fez, kingston & 188 & P1--P4 & fail (ill-conditioned criterion)\\
Q3. Rule on a 33/65-qubit chain & fez & 68 & C1--C3 & pass\\
Q4. Offset vs idle time & fez, marrakesh & 356 & O1--O3 & pass\\
Q5. Idle noise switched by design & kingston, fez, marrakesh & 308 & Z0--S1 & pass on 2; miss 0.001 on 1\\
Q6. Ceiling saturation, $m=3$ & fez & 55 & S1--S3 & pass\\
Q7+8. 2-D rule and one-gate repair & fez & 115 & P1--P3 & pass\\
Q10. Hidden rule, standard template & fez & 80 & R1--R3 & pass\\\bottomrule
\end{tabular}
\end{table}
Runs 1--3b used the first open-plan key.

\paragraph{Simulation.} Statevector and density-matrix simulation in NumPy and PennyLane 0.44. Spectra from FFTs over $2N_{\mathrm{enc}}+3$ equispaced points, averaged over random
parameter draws; a frequency is in the support if its mean power exceeds $10^{-12}$.
\paragraph{Training.} The training recipe of the QNLEP non-linear-capacity stage~\cite{qnlep2026}: 36 training points on $[-\pi,\pi]$, Adam with learning rate 0.1, initialisation $\mathcal N(0,0.1)$, 50 steps (100 for ancilla
re-uploading, 120 for the reservoir), $R^2$ on 120 evaluation points; seeds as stated.
\paragraph{Hardware.} EstimatorV2 with TREX, gate twirling and automatic dynamical decoupling off unless stated; 10 points $x\in\{2\pi j/10\}$ so that $x+\pi$ is on the grid; even part
$\tfrac12(f(x)+f(x+\pi))$. Thresholds, job identifiers, QPU seconds and raw data are in the repository.
The 65-qubit chain was laid along a simple path of the coupling graph, so no routing was needed; its exact ideal follows from simulating the 17-qubit backward light cone of
each readout. Parameters were pre-selected for signal by a rule fixed in advance (the first seeds whose ideal odd power exceeds a threshold).
\paragraph{Real data.} Two features per dataset, min-max scaled to $[-\pi,\pi]$ on the training split, targets standardised; 200 Adam steps (rate 0.05) with a trainable output
scale, five seeds; the structural prediction is a ridge fit ($\lambda=1$) on the model's exact 2-D Fourier support.

\section*{Data availability}
The pre-registrations, same-day predictions, raw hardware results with job identifiers, and all simulation and training outputs that support the findings of this
study are available at\url{https://drive.google.com/file/d/1neEtKzv4DIVoviXKuBWrHzt1PW4j-ham/view?usp=sharing}.

\section*{Code availability}
The code for every simulation, census, training run and hardware experiment is available at \url{https://drive.google.com/file/d/1neEtKzv4DIVoviXKuBWrHzt1PW4j-ham/view?usp=sharing}. A single command
re-scores every hardware experiment reported here against its pre-registered criteria from the stored results, regenerates the figures and builds this
manuscript; a full run also repeats the simulations and the training. The repository's documentation describes both.

\section*{Acknowledgements}
We acknowledge the use of IBM Quantum services for this work. The views expressed are those of the authors, and do not reflect the official policy or position of IBM or
the IBM Quantum team. This work was partially funded by the Institute of Computer Science at Zurich University of Applied Sciences through the project \textnumero\,79838 ``QML -- Quantum Machine Learning''~\cite{zhaw2026qml}.

\end{document}